\documentstyle[prb,aps,twocolumn]{revtex}
\begin{document}
\title{Minimum energy paths for dislocation nucleation in
strained epitaxial layers}
\author{O. Trushin$^1$}
\address{Institute of Microelectronics and Informatics,
Academy of Sciences of Russia, Yaroslavl 150007, Russia}
\author{E. Granato$^2$}
\address{Laborat\'orio Associado de Sensores e Materiais,
Instituto Nacional de Pesquisas Espaciais, 12201--970 S\~ao Jos\'e
dos Camps, SP Brasil}
\author{S.C. Ying}
\address{$^2$Department of Physics, Brown University,
Providence, RI 02912 USA }
\author{P. Salo and T. Ala-Nissila$^2$}
\address{$^1$Helsinki Institute of Physics and Laboratory of Physics,
Helsinki University of Technology, FIN--02015 HUT, Espoo, Finland}

\twocolumn[\hsize\textwidth\columnwidth\hsize\csname@twocolumnfalse\endcsname

\draft

\maketitle

\begin{abstract}
We study numerically the minimum energy path and energy barriers
for dislocation nucleation in a two-dimensional atomistic model of
strained epitaxial layers on a substrate with lattice misfit.
Stress relaxation processes from coherent to incoherent states for
different transition paths are determined using saddle point
search based on a combination of repulsive potential minimization
and the Nudged Elastic Band method. The minimum energy barrier
leading to a final state with a single misfit dislocation
nucleation is determined. A strong tensile-compressive asymmetry
is observed. This asymmetry can be understood in terms of the
qualitatively different transition paths for the tensile and
compressive strains.
\end{abstract}


 ]

\date{May 20, 2002}

The growth and stability of heteroepitaxial overlayers is one of
the central problems in current materials science. Energy-balance
arguments for the competition between strain energy build-up and
strain relief due to dislocation nucleation in mismatched
epitaxial films lead to the concept of an equilibrium critical
thickness. This is defined as the thickness at which the energy of
the epitaxial state is equal to that of a state containing a
single misfit dislocation \cite{bean}. The predicted critical
value from this equilibrium consideration however, both from
continuous elastic models \cite{matthews,vdm} and from models
incorporating layer discreteness \cite{gky}, is much smaller than
the observed experimental value for the breakdown of the epitaxial
state. This suggests that the defect-free (coherent) state above
the equilibrium critical thickness is metastable \cite{tsao} and
the rate of dislocation generation is actually controlled by
kinetic considerations. The idea of strain relaxation as an
activated process is supported by experimental results for the
temperature dependence of the critical thickness
\cite{tsao,zou96}. It is also the fundamental assumption in
kinetic semi-empirical models \cite{hou91}.

Physically, it is expected that the lowest energy barrier for the
nucleation of dislocations would correspond to a path that
initiates from the free surface (with or without defects). Such
processes have been considered in a number of studies within
continuum models \cite{spencer,cullis,grilhe}. It has been pointed
out that surface steps and surface roughness that are not included
in the continuum model could play an important role for
dislocation nucleation \cite{dong98,ich95,tersoff94,brochard}.
Thus, atomistic study is important for a detailed understanding
and direct determination of the mechanisms for defect nucleation
in epitaxial films. However, determination of the correct
transition path and the nucleation barrier from the initial
coherent state to the final state with misfit dislocations is an
extremely challenging problem in an atomistic model.  There are
many saddle points and transition paths in the neighborhood of the
initial coherent state. A brute force molecular dynamics (MD)
study is impractical because of the rare event nature of the
activated processes. In recent years, great progress has been made
in the general field of search for transition paths for
complicated energy landscapes \cite{neb,relax}. In addition,
various accelerated hyperdynamics schemes \cite{hyper1,hyper2}
have been developed to overcome the computational problems for
rare events. However, these schemes still involve considerable
computational efforts for complicated, large energy barriers and
often require a qualitative picture of the energy landscape as a
starting point. Recently we have developed a repulsive potential
minimization method \cite{unp} which allows automatic generation
of many paths leading away from the initial minimum energy
coherent state. When combined with the Nudged Elastic Band method
(NEB) \cite{neb}, this method can be used to efficiently locate
saddle point configurations and barriers for each transition path
without having to make any specific assumptions about the nature
of the transition path.

For the present study, we consider a two-dimensional model of the
epitaxial film and substrate where the atomic layers are confined
to a plane as illustrated in Fig. \ref{patht} (a). Interactions
between atoms in the system are modelled by a generalized
Lennard-Jones pair potential \cite{zhen83}, that is modified to
insure that the potential and its first derivative vanish
\cite{dong98} at a cut-off distance $r_c$ as
\begin{eqnarray}
&&U(r)=  V(r) , \qquad   r \leq r_0;  \cr &&U(r)=V(r) \left[ 3
\left( \frac {r_c-r}{r_c-r_0} \right) ^2 - 2 \left( \frac
{r_c-r}{r_c-r_0} \right) ^3 \right] , \ r>r_0, \label{LJ}
\end{eqnarray}
where
\begin{equation}
V(r)=  \varepsilon \left[ \frac m{n-m} \left( \frac {r_0}r \right)
^n - \frac n{n-m} \left( \frac{r_0}r \right) ^m \right],
\end{equation}
and $r$ is the interatomic distance, $\varepsilon$ the
dissociation energy and $r_0$ the equilibrium distance between the
atoms. This potential has been used previously \cite{dong98}, with
$n=12$ and $m=6$, in a Monte Carlo simulation of epitaxial growth.
We have chosen the value $n=8$ and $m=5$ for the present study. In
contrast to the standard $6-12$ potential, this $5-8$ potential
has a slower fall-off. Thus, when combined with the variation of
the cutoff radius $r_c$, this choice allows us to systematically
study the effect of the range of the potential on misfit
dislocation. Also, the $5-8$ potential gives a more realistic
description of metallic systems than the $6-12$ case. The
equilibrium interatomic distance $r_0$ was set to different values
for the substrate, epitaxial film, and the substrate-film
interfaces. The substrate $r_0=r_{\rm ss}$ and the epitaxial film
$r_0=r_{\rm ff}$ parameters were varied to give a misfit $f$
between lattice parameters as $f=(r_{\rm ff}-r_{\rm ss})/ r_{\rm
ss}$. For the film-substrate interaction we set the equilibrium
distance as the average of the film and substrate lattice
constants, $r_0=r_{\rm fs}=(r_{\rm ff}+r_{\rm ss})/2$. Positive
misfit $f$ corresponds to compressive strain and negative $f$ to
tensile strain when the film is coherent with the substrate.
Calculations were performed with periodic boundary conditions in
the direction parallel to the film-substrate interface. Typically,
one-dimensional layers containing 50 atoms or more were used in
the calculations. In the calculations the bottom five layers
represented the substrate, with the last two layers held fixed to
simulate a semi-infinite substrate while all other layers were
free to move.

Our new scheme of identifying the saddle points and the transition
paths consists of several stages. First, the initial epitaxial
state is prepared by minimizing the total energy of the system
using MD cooling. This leads to an initial coherent epitaxial
state in which the interlayer spacing is relaxed, but the atoms
within the layers are under compressive or tensile strain
according to the misfit. Next, we introduce a short-ranged
repulsive potential centered at the starting epitaxial
configuration of the form
\begin{equation}
U_{tot}(r)=U(r)+A\exp\{-\alpha(r-r_{0})^2\},
\end{equation}
where $r_{0}$'s are the coordinates of the initial state at the
minimum \cite{relax}. The basic idea here is to modify the local
energy surface to make the initial epitaxial state unstable. When
the system is slightly displaced from the initial state (randomly
or in a selective way), it will then be forced to move to nearby
minimum energy states. By making this repulsive potential
sufficiently localized around the initial potential minima, the
surrounding minima would be unaffected and the final state energy
would then depend only on the true potential of the system. This
method can generate many different final states depending on both
the initial displacements and the parameters of the repulsive
potential.  In this work, we only consider final configurations
corresponding to a single misfit dislocation. Rather than trying
random initial displacements, some knowledge of the dislocation
generation mechanism is useful for expediting the process. Given
the knowledge of the final state, we then use the NEB method to
locate the saddle point and find the minimum energy path (MEP)
between the initial and final states. We note that the path
generated in the first part of moving away from the repulsive
potential can be used as an initial guess for the MEP
determination in the NEB method.

We have performed calculations for various misfits but in this
paper we concentrate on the case of a relatively large 8\% misfit.
We work with dimensionless quantities with $\varepsilon$ as the
energy scale and $r_{\rm ss}$ as the length scale. Two different
choices of cutoff, namely $r_c$=1.5 $r_{\rm ss}$ and $r_c$=4.7
$r_{\rm ss}$ were made. The results for the barriers are
qualitatively similar, so we present here only the results for the
short range potential with $r_c$=1.5 $r_{\rm ss}$. However, the
equilibrium critical thickness and its asymmetry with respect to
tensile or compressive strain are very sensitive to the range of
the potential \cite{unp}.

The results for the MEP from coherent to incoherent states are
shown in Fig. \ref{patht} for a film under tensile strain and Fig.
\ref{pathc} for compressive strain. They show clearly the
existence of an energy barrier for the nucleation of a misfit
dislocation. Thus, the non-equilibrium critical thickness can be
much larger than the equilibrium value and it is controlled in
practice by the kinetics of defect nucleation.

For compressive strain, the final state is characterized by the
presence of an adatom island on the surface of the film for each
misfit dislocation. The number of adatoms in the island exactly
corresponds to the number of layers in the film. Such form of the
final state is determined by the geometry of the misfit
dislocation. For every misfit dislocation, an extra atom is
removed from each layer to relieve the compressive stress.  For
tensile strain, the final state is characterized by the presence
of pits on the surface. Again, the size of the pit is determined
by geometrical considerations. For every misfit dislocation, an
extra atom has to be added to each layer to relieve the tensile
stress. For both cases, the dislocation core is localized in the
substrate-film interface region.

Figures \ref{patht} and \ref{pathc} also show the particle
configurations at the different points along the MEP which reveal
details of defect nucleation and strain relaxation process. The
transition path for the compressive strain has a more local
nature, with relatively fewer bonds involved initially, whereas
for the tensile strained film, the nucleation proceeds via a more
collective path, involving concerted motion along glide planes.
The energy barrier for nucleation of a dislocation is much higher
for the compressive strain relative to the case of tensile strain.
This asymmetry is very robust and it persists when we change the
range of the potential by varying the cutoff.

To understand the origin of this asymmetry, we plot in Fig.
\ref{dist} the distribution of the nearest-neighbor bond lengths
for the film from the initial epitaxial film to the saddle point
configuration for both the compressive and tensile cases. It can
be seen that the behavior of the compressively strained film and
the tensile-strained film is very different. In the tensile case,
the redistribution of the bond lengths going from the initial
coherent state to the saddle point configuration involves a
significant contraction of the intralayer bonds leading to partial
relaxation of the tensile strain in the film. On the other hand,
for the compressively strained film, the initial delta function
peak for the intralayer bond lengths broadens almost symmetrically
and  there are no significant relaxation of the compressive strain
in the film. This explains the relatively higher energy costs and
a corresponding larger nucleation barrier for the compressive
strained film.  Microscopically, the origin of the different
behavior could arise from the strong anharmonicity of the
interaction potential. For the compressive strain, intralayer
rearrangements involve some further compression of the bonds which
is energetically costly. Thus, a more localized initial
configuration with a higher barrier results as opposed to the
collective behavior of the tensile strained layer. We have also
checked that the boundary conditions and system sizes do not
affect the results qualitatively by comparing results from systems
with periodic and free boundary conditions, and for layers twice
as long.

In summary, we have developed a new scheme of identifying minimal
energy path for spontaneous generation of misfit dislocation in an
epitaxial film. This new approach requires no a priori assumptions
about the nature of the  transition path or the final states. A
nonzero activation barrier for dislocation nucleation is found in
the minimum energy path from coherent to incoherent state above
the equilibrium critical thickness, confirming the meta-stability
of the epitaxial coherent film. The nucleation mechanism from a
flat surface depends crucially on whether we start from a tensile
or compressive initial state of the film. This asymmetry
originates from the anharmonicity of the interaction potentials
which leads to qualitatively different transition paths for the
two types of strains.  A tensile-compressive asymmetry has also
been found previously \cite{dong98,ich95} in other contexts. The
present method can be extended to three-dimensional models with
more realistic interaction potentials. Preliminary calculations
for the Pd/Cu and Cu/Pd systems \cite{unp} with the Embedded Atom
Model potentials \cite{eam} confirms the effectiveness of the
method in three dimensions. These results will be published
elsewhere.

\bigskip

This work was supported by a NSF-CNPq grant (E.G. and  S.C.Y.), by
the Russian Ministry of Science and Technology (O.T.), FAPESP
(E.G.), and by the Academy of Finland through its Center of
Excellence program (O.T., P.S., and T.A-N.).

\begin{figure}
\caption{Particle configurations and energy change $E_i-E_0$ at
different states (images) along the minimum energy path, for
tensile strain ($f=-8\%$). The two layers at the bottom are held
fixed while all others are free to move. Open circles represent
the substrate atoms and filled circles the epitaxial film. Only
the central part of the layers with major atom rearrangements is
shown.} \label{patht}
\end{figure}

\begin{figure}
\caption{Same as Fig. \ref{patht} but for compressive strain ($f=+8\%
$).} \label{pathc}
\end{figure}

\begin{figure}
\caption{Nearest-neighbor bond distributions of the epitaxial film
at the saddle point for the (a) tensile, and (b) compressive
cases. Solid and dotted arrows indicate the position of the
delta-function peak corresponding to intralayer and interlayer
bond distributions of the initial coherent film. } \label{dist}
\end{figure}

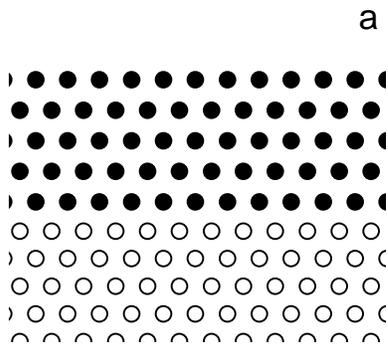
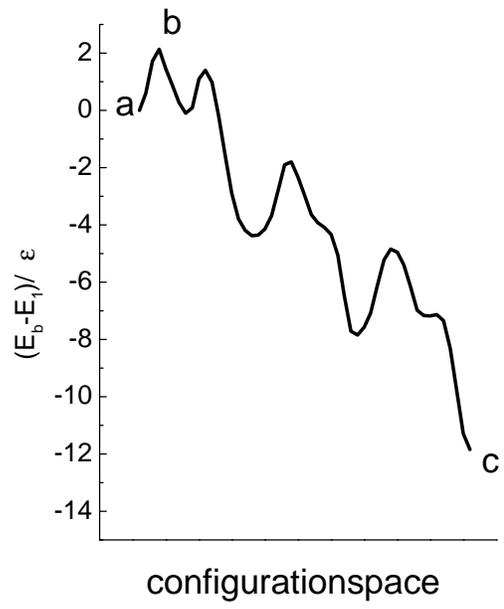
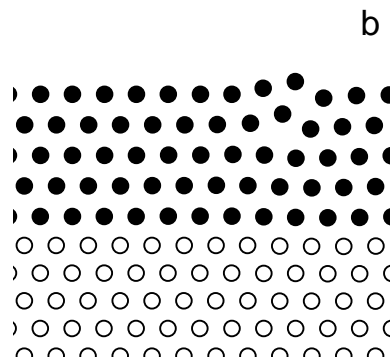
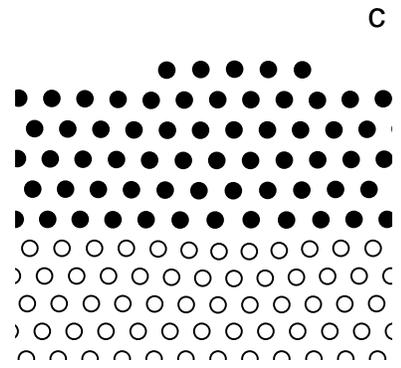
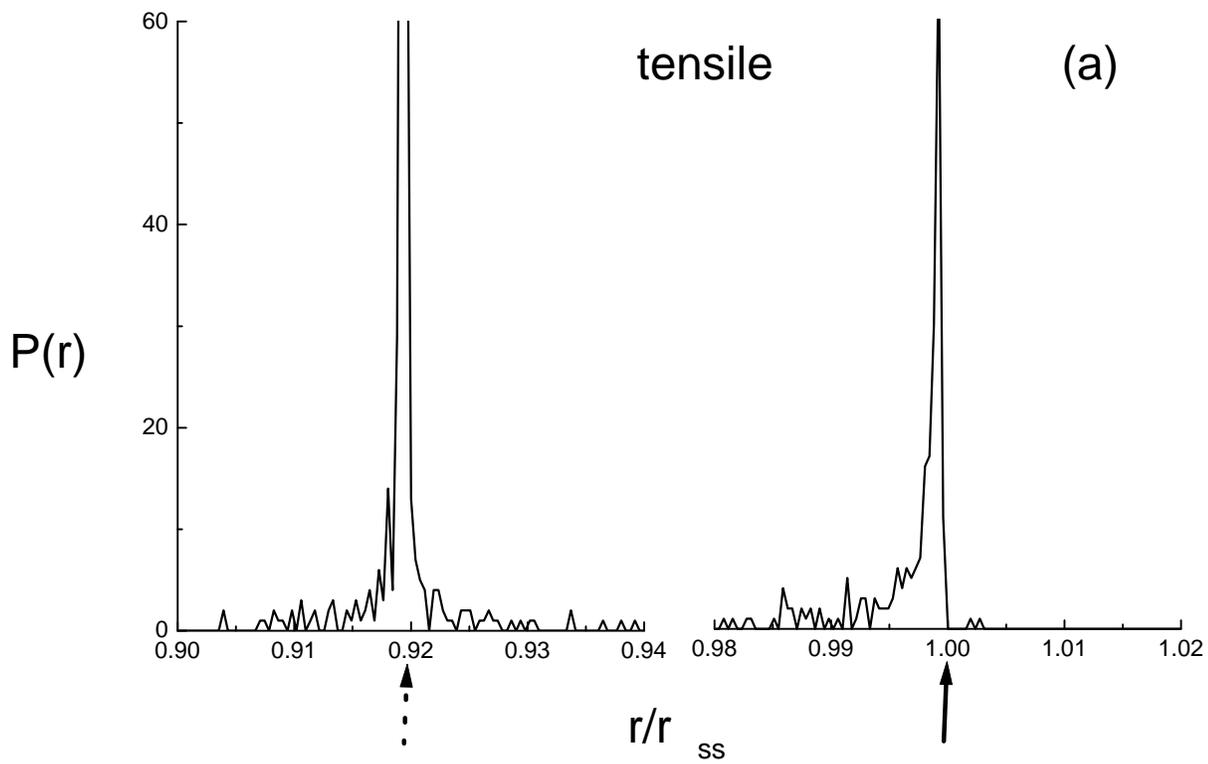
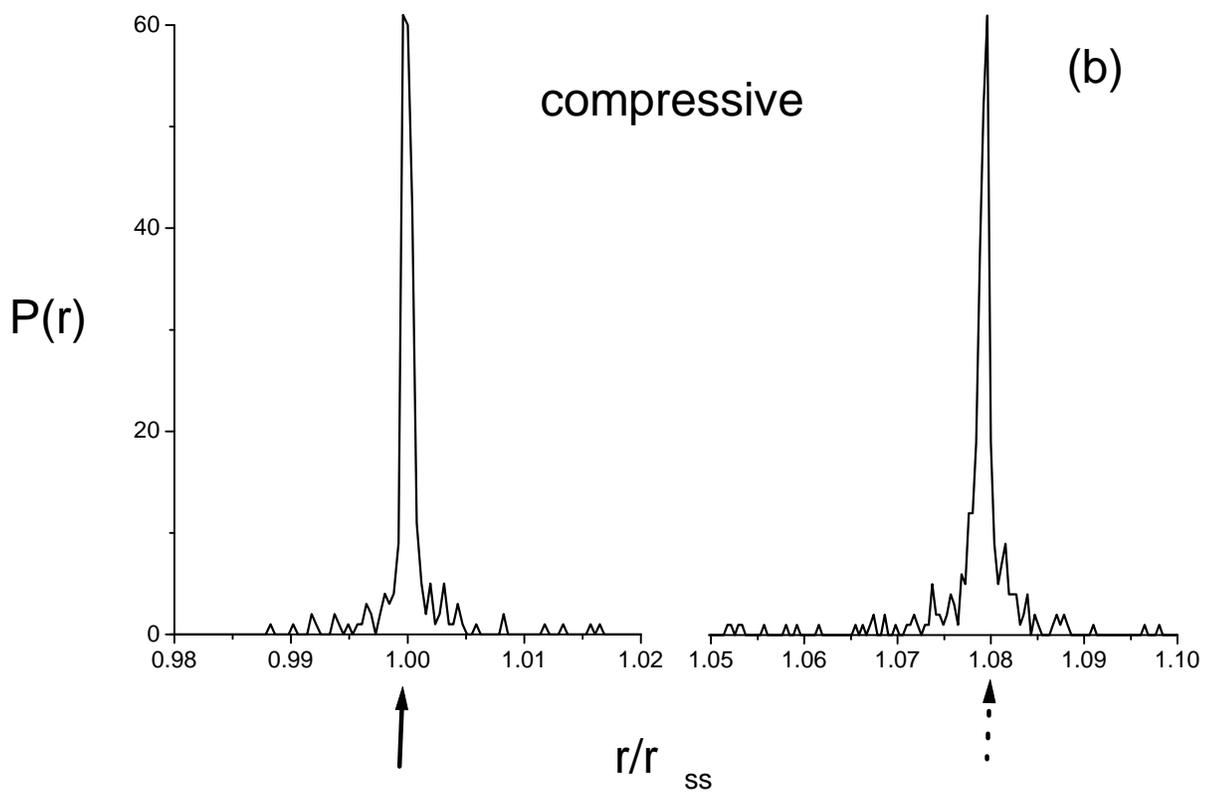
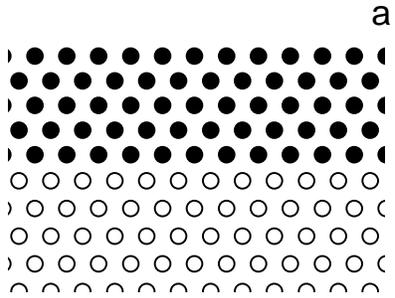
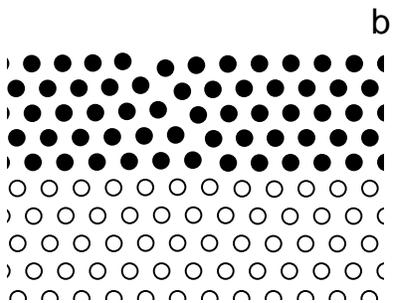
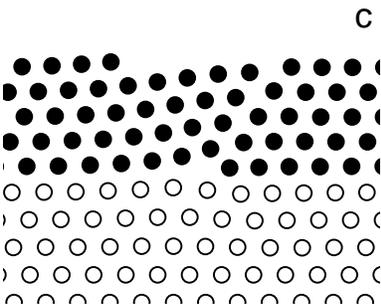
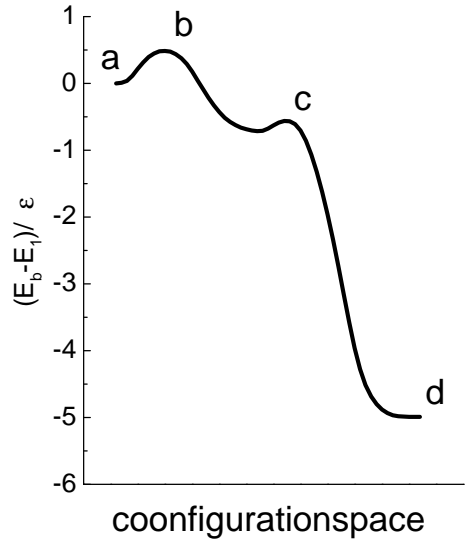
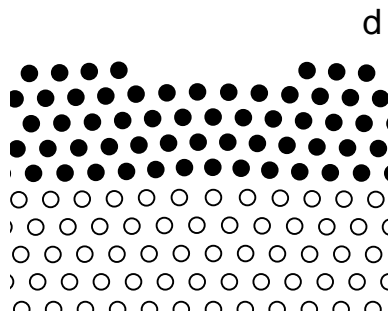
\end{document}